\documentclass[
    a4paper,
    prb,
    bibnotes,
    twocolumn,
    showpacs,
    superscriptaddress,
    groupedaddress
]{revtex4}

\usepackage{array}
\usepackage{tikz}
\usepackage{graphicx}
\usepackage{amssymb}
\usepackage{amsmath}
\usepackage{fullpage}
\usepackage{hyperref}
\usepackage{paralist}
\usepackage{subfigure}
\usepackage{mathdots}
\usepackage{float}
\hypersetup{
    colorlinks   = true,
     citecolor    = blue,
     linkcolor    = blue,
     citecolor    = blue,
     filecolor     = blue,
     urlcolor	      =	blue  
}

\newcommand*{\citen}{}
\DeclareRobustCommand*{\citen}[1]{%
  \begingroup
    \romannumeral-`\x 
    \setcitestyle{numbers}%
    \cite{#1}%
  \endgroup
}

\setcitestyle{square}

\newcommand{\beq}{\begin{equation}}
\newcommand{\eeq}{\end{equation}}
\newcommand{\bra}[1]{\langle #1 |}
\newcommand{\ket}[1]{| #1 \rangle}
\newcommand{\brahket}[3]{\langle #1 | #2 | #3 \rangle}

\newcommand{\ketbra}[2]{\ket{#1}\bra{#2}}

\newcommand{\mybpar}[1]{\left( #1 \right)}

\newcommand{\beqa}{\begin{eqnarray}}
\newcommand{\eeqa}{\end{eqnarray}}

\newcommand{\etal}{\mbox{\textit{et al.}}}

\newcommand{\Eqref}[1]{Eq.~(\ref{#1})}
\newcommand{\Figref}[1]{Fig.~\ref{#1}}
\newcommand{\Secref}[1]{Sec.~\ref{#1}}
\newcommand{\Tabref}[1]{Tab.~\ref{#1}}

\begin{document}
\title{Theory of vibrationally assisted tunneling for hydroxyl monomer flipping on Cu(110)}
\date{\today}

\author{Alexander \surname{Gustafsson}}
\email{alexander.gustafsson@lnu.se}
\affiliation{Department of Physics, Linnaeus University, 391 82 Kalmar, Sweden}
\author{Hiromu \surname{Ueba}}
\email{ueba@eng.u-toyama.ac.jp}
\affiliation{Division of Nano and New Functional Materials Science, Graduate School of Science and Engineering, University of Toyama, Toyama, Japan}
\author{Magnus \surname{Paulsson}}
\email{magnus.paulsson@lnu.se}
\affiliation{Department of Physics, Linnaeus University, V\"axj\"o, Sweden}

\begin{abstract}
To describe vibrationally mediated configuration changes of adsorbates on surfaces we have developed a new theory to calculate both reaction rates and pathways. The method uses the T-matrix to describe excitations of vibrational states by the electrons of the substrate, adsorbate and tunneling electrons from a scanning tunneling probe. In addition to reaction rates, the theory also provides the reaction pathways by going beyond the harmonic approximation and using the full potential energy surface of the adsorbate which contains local minima corresponding to the adsorbates different configurations.
To describe the theory, we reproduce the experimental results in [T. Kumagai \etal, Phys. Rev. B \textbf{79}, 035423 (2009)], where the hydrogen/deuterium atom of an adsorbed hydroxyl (OH/OD) exhibits back and forth flipping between two equivalent configurations on a Cu(110) surface at $T = 6$ K. We estimate the potential energy surface and the reaction barrier, $\sim$160 meV,  from DFT calculations. The calculated flipping processes arise from i) at low bias, tunneling of the hydrogen through the barrier, ii) intermediate bias, tunneling electrons excite the vibrations increasing the reaction rate although over the barrier processes are rare, and iii) higher bias, overtone excitations increase the reaction rate further.

\end{abstract}


\maketitle

\section{INTRODUCTION}
Electron transport through single-molecule junctions has been receiving broad of interest for a development of novel molecular devices. Nonlinear current-voltage characteristics associated with the vibrationally mediated configurational change with different conductance have been observed in pyrrolidine on a Cu(001),\cite{Gaudioso2000} CO-Pt junction,\cite{Thijssen2006} and for a flipping motion of hydroxyl dimer on Cu(110).\cite{Kumagai2009, Ootsuka2011, Okuyama2012} In Ref.~\citen{Kumagai2009} Kumagai \etal~  also reported that the inclined OH monomer axis switches back and forth between two equivalent orientations via hydrogen-atom tunneling. The motion is enhanced by tunneling electron that excites the OH bending mode that directly correlates with the reaction coordinate. This experimental work motivated Davidson \etal\cite{Davidson2010} to explore the quantum nuclear tunneling dynamics of hydroxyl on Cu(110) using DFT based techniques. They calculated the flip rates by tunneling in the vibrationally ground state for two-dimensional (as a function of oxygen and hydrogen displacement) potential energy surface. The potential barrier for the flipping of OD monomer is estimated to 140-180 meV by the DFT calculations of the transition path.\cite{Kumagai2009,Davidson2010}
 
The fractional occupation of the high-current state for OD measured below about $V=40$ mV seems to suggest quantum tunneling of a deuterium atom in the vibrational ground state. The flip motion is enhanced and the occupation for the high conductance (HC) state increases with an increase in $V$. Considering a potential barrier separating high and low conductance states associated with a reversible flip motion of H(D) atom anchored by O atom bound to a Cu substrate atom, the flip motion of OD dimer on Cu(110) does not occur by an over-barrier process through incoherent multi-electron vibrational ladder climbing. It is certainly induced by a vibrational assisted tunneling (VAT) in both the ground and excited states. A theory of VAT coupled together with the incoherent vibrational ladder climbing has been developed by Tikhodeev and Ueba.\cite{Tikhodeev2009} This multi-level VAT has been taken into account for a Xe atom transfer,\cite{Walkup1993} and for hydrogen-bond exchange within a single water dimer on Cu(110).\cite{Kumagai2008}
 In this work we propose a novel method based on the T-matrix to calculate VAT transition rate between a double well potential beyond harmonic approximation. This allows us to take into account forward and backward tunneling, both in the ground and excited states between different vibrational levels in each potential. We calculate the tunneling transition rates and path of reaction for various bias voltages at a fixed tunneling current. Combined with the inelastic electron tunneling spectrum (IETS\cite{Persson1987,Ho2002,Paulsson2008}) it is clarified how the VAT manifests itself in the flipping rates as a function of bias voltage and tunneling current. The anharmonic potential causes an overtone signal having comparable intensity to the fundamental peak in the IETS. This gives rise to the corresponding threshold in the $R(V)$ curve. Our results also underline that the $R(I)$ curve exhibits a crossover from a single electron process to two electron process with increasing tunneling current. 
 
\section{THEORY}
To develop a theory capable of describing reactions at surfaces we will start
from a non-harmonic potential energy surface, i.e., below we will use a double well potential surface. 
From the exact diagonalization of the non-interacting vibrational system we develop a rate equation describing the transitions between vibrational states and thus the reaction rate.
The transition rates caused by tunneling electrons are calculated from the electron-phonon (\textit{e}-ph) interaction using the impulsive approximation of the T-matrix.\cite{Persson2009} From the theory developed here it is possible to include the effects of the non-harmonic potential together with \textit{e}-ph interactions to arbitrary order.
To illustrate the theory, we apply it to the switching of OD monomers using DFT to obtain the potential landscape, \textit{e}-ph coupling and electronic structure. 

\subsection{Potential landscape}
\begin{figure}[tbh!]
\begin{tikzpicture}
    \node[anchor=south west,inner sep=0] at (.275,0) {\includegraphics{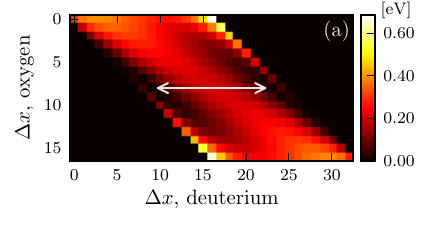}};
    \node[anchor=south west,inner sep=0] at (0,-4.6) {\includegraphics{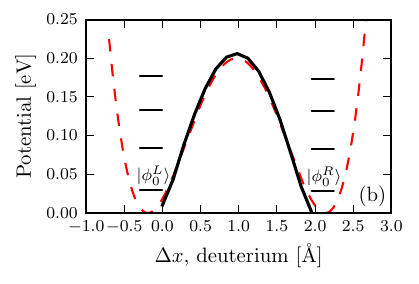}};
    \node[anchor=south west,inner sep=0] at (3.6,-3.3) {\includegraphics[scale=.035]{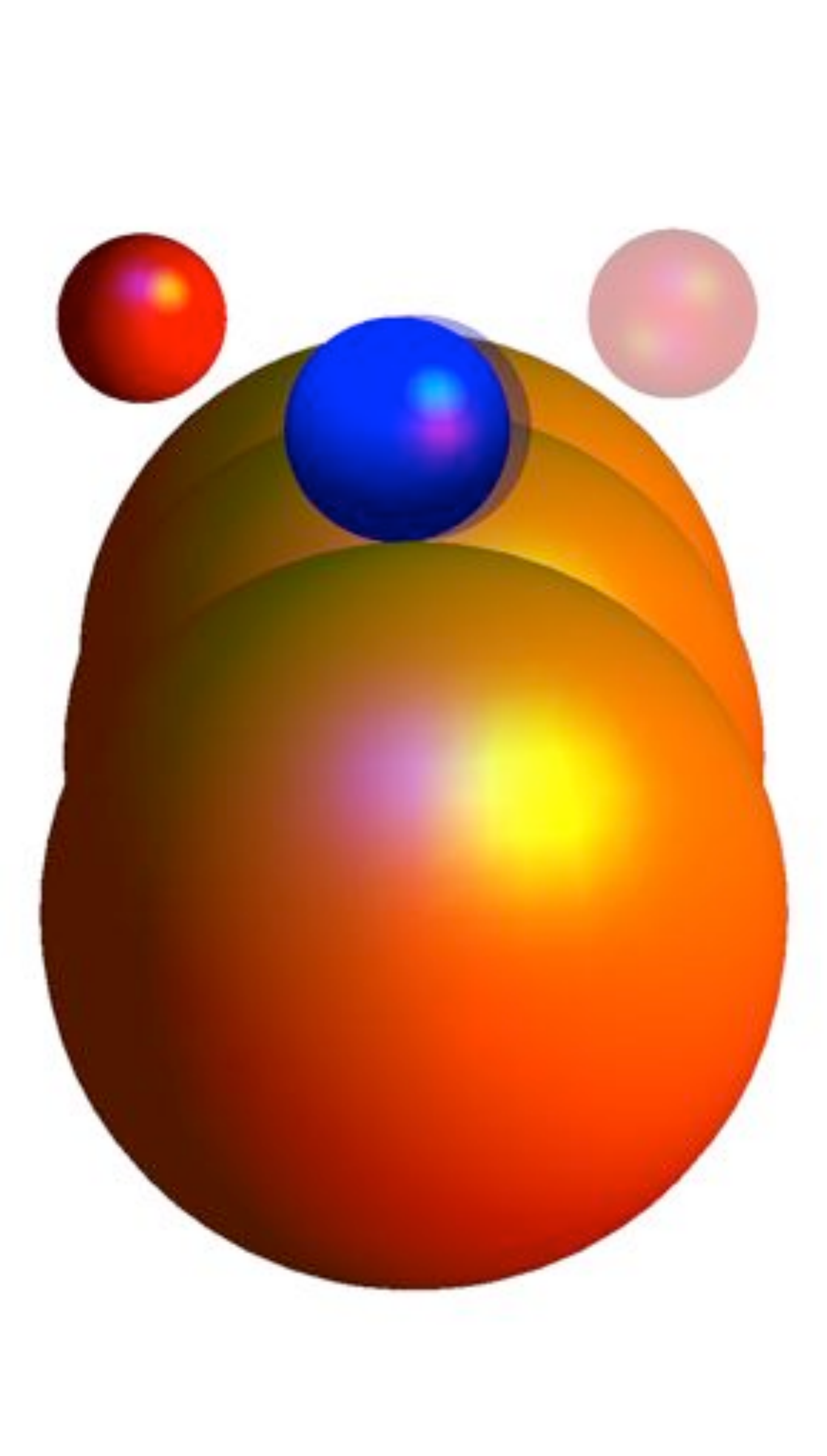}};
    \draw[ thin,stealth-stealth] (4.3,-2.0) arc (30:150:.32) ;
\end{tikzpicture}
	\caption{(a) Potential energy surface as a function of $x$-position of the O and D atoms. (b) Potential energy barrier keeping the O atom fixed. Solid black corresponds to white arrow in (a), and dashed red line showing the shape of the quartic function used as a 1D potential in the numerical simulation\label{PES}}	
\label{fig.PES}
\end{figure}

To facilitate the presentation of the theory we exemplify the procedure using the potential landscape for the OD monomer on a Cu(110) surface, see Fig.~\ref{PES}(a), where computational parameters are given in the next section. The potential depends on the motion of both the oxygen and deuterium atoms along the surface while the height is allowed to relax. Although the minimal energy path deviates slightly from a straight path, we approximate the reaction with a 1D potential along the motion of the deuterium atom alone, see Fig.~\ref{PES}(b). We further introduce an asymmetry between the wells to make the eigenfunctions localized. According to the experiment, the magnitude of the asymmetry depends on the STM tip position, which is explained more carefully in \Secref{Results}.

Solving the Schr\"odinger equation numerically for the deuterium motion in the potential  provides us with the exact vibrational states $\ket{\phi_{\text{ph}}}$. The low lying states correspond well with the harmonic approximation while the higher energy states are delocalized above the barrier. The barrier height $\sim160$ meV is approximately three times the vibrational energy quanta in the harmonic approximation. At low bias (below fundamental frequency) reactions are unlikely, apart from direct tunneling between the vibrational ground states.

\subsection{Transition rates}\label{FGRTmatrix}
Our starting point to obtain the transition rates is the known vibronic states together with the electronic states and the \textit{e}-ph interaction $\bold{M}$. The \textit{e}-ph coupling is obtained by expanding the electronic Hamiltonian to lowest order around the equilibrium geometry (see Ref.~\citen{Frederiksen2007} for details) 
\beq
\hat{H}=\hat{H}_e^0+\hat{H}_{e-\text{ph}}.
\eeq
where $H_0$ is the electron Hamiltonian of a static lattice, and where the perturbation reads
\beq
H_{e-\text{ph}}=\widetilde{\bold{M}} \hat{x}=\sum_{\alpha\beta}\bold{M}_{\alpha\beta}c_\alpha^{\dagger}c_\beta (a^{\dagger}+a),
\eeq
where $\hat{x}$ is the vibration coordinate operator which, in the harmonic approximation, is associated with the creation/destruction operators $a^\dagger, a$, and $\bold{M}=\sqrt{\frac{\hbar}{2 m \omega_{\lambda}}}\widetilde{\bold{M}}$  the normal \textit{e}-ph interaction matrix of vibrational mode $\lambda$. 

The continuum of electronic scattering states ($\ket{\psi_{\alpha}}$), where the index $\alpha$ can for instance be composed of the energy ($E^\alpha_e$) and an integer for the band index, are scattering states for the infinite electrode-device-electrode system ($\hat{H}_e^0$). We further divide the states into scattering states originating in the substrate/tip contact.\cite{Paulsson2007} In our case, the interaction and electronic states are calculated from DFT using the non-equilibrium Green's function techniques (described below).  

We then form the direct product of the electronic and vibrational one particle states $\ket{\Psi_{i,\alpha}}=\ket{\psi_{\alpha}} \otimes \ket{\phi_{i}}$ with the total energy $E^{i,\alpha}=E_e^\alpha+E_{\text{ph}}^i$ (electron+vibration). The transition rates $w^{\alpha\to\beta}_{i\to f}$ from initial ($i$ vibrational and $\alpha$ electronic states) to final ($f$, $\beta$) state is calculated using the T-matrix approach ($i,\alpha \ne f,\beta$),
\beq
w^{\alpha\to\beta}_{i\to f}=\frac{4\pi}{\hbar}\left|\bra{\Psi_{f,\beta}}\bold{T}\ket{\Psi_{i,\alpha}}\right|^2\delta(E^{i,\alpha}-E^{f,\beta}),\label{FGR}
\eeq
where $\bold{T}=\bold{M}+\bold{M}\bold{G}\bold{M}$, and $\bold{G}$ is the exact Green's function for the combined electron-phonon state. We underline that this expression is exact, although in practice approximations to $\bold{G}$ are usually made. We further notice that the extension to higher order in $\bold{M}$ is tractable within the impulsive approximation. \footnote{$\bold{G}\approx \bold{G}^e \otimes \bold{I}$, i.e., the slow evolution of the vibrational degree of freedom is ignored during the scattering event.} Higher order terms, e.g., expanding the T-matrix to second order, become important when the bias voltage exceeds two vibrational quanta, which will be discussed further in the results section.
In this paper we will limit ourself to expansion to lowest order in the T-matrix, $\bold{T}\approx\bold{M}$. 
 
The energy resolved rates can be simplified by using the spectral functions for the substrate/tip $\bold{A}_{a}(E_e^{a})={2 \pi} \sum_j\ketbra{\psi_j}{\psi_j}$ which contains the electronic states from lead $a$ (substrate or tip) at energy $E_e^a$.\cite{Paulsson2007} Summing over the electronic 
scattering states originating from lead $a$ ($b$),
\begin{widetext}
\beqa
w^{a\to b}_{i\to f} (E_e^a)&=&
\frac{4\pi}{\hbar} 
	\sum_{\alpha,\beta\in a,b} 
		{\left| \brahket{\psi_\beta(E_e^\beta)}{\widetilde{\bold{M}} }{\psi_\alpha(E_e^\alpha)}
		\right|^2 \left| \brahket{\phi_f}{\hat{x}}{\phi_i}\right|^2}
	\delta(\Delta E)
\\
&=& 
\frac{1}{\pi\hbar}
\underbrace{
	\text{Tr}[\widetilde{\bold{M}} \bold{A}_b (E_e^a-\Delta E_{\text{ph}}) \widetilde{\bold{M}}\bold{A}_a (E_e^a)]}_
	{\text{electron part}}
\underbrace{
	\left |\brahket{\phi_f}{\hat{x}}{\phi_i}\right|^2}_
	{\text{vibrational part}}
\eeqa
\end{widetext}
where $\Delta E=E^{i,a}-E^{f,b}$, $\Delta E_{\text{ph}}=E_{\text{ph}}^{f}-E_{\text{ph}}^{i}$.
The scattering rate $w^{a\to b}_{i\to f}(E^a_e)$ thus provides the rate of scattering between initial (\textit{i}) and final (\textit{f}) vibrational states involving electrons in initial(final) states originating from lead \textit{a}(\textit{b}) at energy $E^a_e$ ($E^b_e = E^a_e -\Delta E_{\text{ph}}$ from energy conservation).

Integrating the energy resolve scattering rates over energy, introducing the Fermi-functions ($f_{a,b}$) of the electrons in the leads to account for the Pauli principle, and assuming that the spectral functions are energy independent close to the Fermi energy gives
\begin{widetext}
\beq
\Gamma_{i\to f}^{a\to b}=
	\int{f_a(E_e^a) \mybpar{1-f_b(E_e^a-\Delta E_{\text{ph}})} w_{i\to f}^{a\to b} (E_e^a) \mbox{d}E_e^a} 
\approx
	\frac{1}{\pi\hbar}
	\text{Tr}\left[\widetilde{\bold{M}}\bold{A}_a\widetilde{\bold{M}}\bold{A}_b\right]
	\left|\brahket{ \phi_f}{\hat{x}} {\phi_i} \right|^2
 	\times \mathcal{F}
	\label{eq.mainRate}
\eeq
\end{widetext}
where
\beqa
\mathcal{F}(\mu_a,\mu_b,\Delta E_{\text{ph}})&=&\frac{eV-\Delta E_{\text{ph}}}{1-\exp\left[-\frac{eV-\Delta E_{\text{ph}}}{kT}\right]}\\
&\underset{kT\to 0} \approx& \Theta(eV-\Delta E_{\text{ph}}) \mybpar{eV-\Delta E_{\text{ph}}},\nonumber 
\eeqa
the applied bias $eV = \mu_{b}-\mu_{a}$ is given by the difference in chemical potentials for the contacts, and $T$ is the temperature. 

For a given transition between vibrational states ($i\to f$) we obtain four rates with respect to the the origin and destination ($a\to b$) of the electron participating in the process. Damping of the vibrational system ($\Delta E_{\text{ph}}<0$) occurs for all four terms although for STM this electron-hole (\textit{e-h}) pair damping is usually largest for the substrate-substrate term since the electronic coupling is larger for a metallic substrate. Excitation of the vibrational system ($\Delta E_{\text{ph}}>0$) only occurs for the tip-substrate term with the correct bias (at $kT\ll \Delta E_{\text{ph}}$), as expected from the Pauli principle.  

To compare with previous results it is useful to consider the harmonic approximation where, by symmetry, only transitions differing by one harmonic quanta is allowed in Eq.~\ref{eq.mainRate}. Furthermore, summing the rates over excitations and relaxations to obtain the total power deposited into the vibronic system recovers the expressions previously obtained using the more rigorous many-body NEGF technique.\cite{Paulsson2005} However, since our potential serves as a perturbed harmonic potential, multi-phonon transitions are possible, though normally with a much smaller probability. We also emphasize that the present theory can be extended to include higher order transitions which can be important to describe energy transfer between vibrational states.
\subsection{Rate equation}
To obtain the reaction rates, the transition rates involved between the vibronic states are calculated by the scheme presented in the preceding section, and used in a conventional rate equation,
\beq
\dot{\vec{n}}(t)=\bold{\Gamma}\cdot \vec{n}(t),
\eeq
where the matrix elements $\Gamma_{ij}=\sum_{a,b}\Gamma^{a\to b}_{i\to j}$ and $\Gamma_{ii}=-\sum_{i\neq j}\Gamma_{ij}$. The steady state solution of the rate equation gives the occupations of the vibrational states,
\beq
 \vec{n}=\{\underbrace{n_1,n_2,...,n_{N/2}}_{n_H},\underbrace{...,n_{N-1},n_N}_{n_{L}}\},
 \eeq
where $N$ is the number of vibrational states taken into consideration in the calculations, and $n_{H,L}$ are the high/low conductance occupations. Since we are investigating the case when bias voltage is well below the reaction barrier at low temperature we use $N=8$ in the simulation, giving four localized vibrational modes on each side of the reaction barrier. The occupations essentially depend on the bias voltage, the tunneling current, electron-phonon coupling, and vibrational states. From the occupations we calculate the time average of the current,\cite{Ootsuka2011}
\beq
I(V)=G_0 (n_H(V)T_H+n_L(V)T_L) V, \label{d2idv2}
\eeq 
where $T_{H,L}$ are the transmission coefficients of the high/low conductance states, and $G_0=e^2/(\pi \hbar)$ is the spin-degenerate conductance quantum.

In order to elucidate the reaction rate and pathway we artificially introduce a source in the initial vibrational state  (left ground state) and remove probability when arriving to the final state (right ground state). The leakage is simulated by magnifying the last diagonal element of $\bold{\Gamma}$, $\Gamma_{ff}=-\sum_{f\neq i}\Gamma_{if}$, by several orders of magnitude. This implies that (almost) everything that arrives to the final state leaves the system, and gives a unidirectional left to right reaction pathway. Injecting one particle per second in $L_0$, i.e., injection rate $\vec{R}^{\text{in}}=\{1,0,0,...\}$, then gives the slightly modified rate equation,
\beq
\widetilde{\bold{\Gamma}}\vec{n}(t)+\vec{R}^{\text{in}}=\dot{\vec{n}}(t),
\eeq
where $\widetilde{\bold{\Gamma}}$ is the previously modified rate matrix. As an approximation we use the solution in the long time limit, i.e., $\dot{\vec{n}}(t)=0$, which gives equilibrium occupations
\beq
\vec{n}=-\widetilde{\bold{\Gamma}}^{-1}\cdot \vec{R}^{\text{in}}.
\eeq 
The reaction rate is obtained as the ratio of the injection rate into $L_0$ and its corresponding equilibrium occupation,
\beq
R(I,V)=\vec{R}^{\text{in}}_0/\vec{n}_0.\label{RateLR}
\eeq

The non-equilibrium rate matrix and occupations are further used to obtain a unidirectional reaction path matrix, calculated by $\widetilde{\bold{\Gamma}}_{ij}\cdot\vec{n}_j/\vec{n}_0$, with diagonal elements set to zero by hand in order to eliminate transitions within the same vibrational state. These matrix  elements are translated to arrows with corresponding magnitudes of individual transition rates between the vibrational states in the next section, see \Figref{PathFig}. This provides a physical picture of the main reaction pathway at various bias voltages and tunneling currents. 

\subsection{Computational details}
The numerical calculations of the potential landscape, electronic states and \textit{e}-ph coupling were performed with the \begin{scriptsize}SIESTA\end{scriptsize}\cite{Soler2002} DFT package. The computations were performed with the PBE GGA functional\cite{Perdew1996} with SZP (DZP) basis set for the copper substrate (molecule), a 400 Ry cutoff energy for the real space grid integration, and $4\times3\times1$ $k$-point sampling. The range of the basis orbitals for D, O and Cu are 3.7, 3.0 and 4.3 \AA\hspace{0pt} respectively.

The Cu(110) substrate is modeled by a six-layer slab with a $3\times3$ periodicity (lattice constant $a=3.64$ \AA). The adsorbate and two topmost Cu layers were relaxed until the residual forces were less than 0.02 eV/\AA. The tip part is modeled by a fixed three-layer slab with periodicity $3\times3$, and one protruding relaxed copper atom, representing the tip. As the experiment is performed with the tip held slightly asymmetrically over the transition state, we place the protruding tip atom over the left configuration, which is assumed to be the initial (HC) configuration throughout.  The elastic electronic transport calculations uses the \begin{scriptsize}{TRANSIESTA}\end{scriptsize}\cite{Brandbyge2002} module, where extra layers are added to give eight(seven) layers in substrate(tip). \begin{scriptsize}{INELASTICA}\end{scriptsize}\cite{Frederiksen2007} uses the elastic transport quantities combined with a vibrational analysis to find the electronic spectral functions and \textit{e}-ph couplings (calculated with displacement $\pm0.02$ \AA\hspace{3pt}of the dynamic atoms in each spatial direction) used to calculate the inelastic scattering rates $\propto\text{Tr}[\widetilde{\bold{M}}\bold{A}_a\widetilde{\bold{M}}\bold{A}_b]$. Along the reaction coordinate, i.e., vibrational mode rot$_z$, this gives a calculated \textit{e-h} pair damping rate ($a=b=$ substrate) $2.4\times10^{11}$ s$^{-1}$, and excitation rate ($a=$ tip, $b=$ substrate) $1.8\times10^9$ (sV)$^{-1}$ at tunneling current 5 nA in HC configuration.

The converged geometries\footnote{We found a second minima with OD bonding angle $73^{\circ}$, and 0.18 \AA\hspace{0pt} lateral oxygen displacement, which we assume is a more unrealistic geometry due to the significantly larger oxygen displacement} give inclined OD bonding angles $65^{\circ}$ to the substrate surface normal, a 0.03 \AA \ lateral oxygen displacement from the short bridge site, and OD bonding length 0.99 \AA. Hence the circular motion between the minima gives an arc length $2.2$ \AA. The choice of tip-position gives high/low conductance ratio $\sigma_H/\sigma_L=1.44$, compared to the experimental value $\sim$1.35, which might suggest that a larger tip-sample distance is used in the experiment. Since the \textit{e}-ph coupling differs between the high- and low conductance states, we choose the one for which the integral $|\bra{\phi_{\text{ph}}^f}\hat{x}\ket{\phi_{\text{ph}}^i}|^2$ has its weight. However, using the same \textit{e}-ph coupling in both wells has a negligible impact on the final results. We only calculate the inelastic scattering rates at one tip-height. To model opening/closing of the tip-sample distance we scale the spectral function of the tip, which affects the elastic conductance and the transition due to electrons in the tip. We assume that this simple approximation works well owing to the large tip-molecule distance, which is 4.7 (5.4) \AA\hspace{0pt} for the left (right) geometry, since only the outermost atomic orbital of the tip atom plays a role.

\section{Results}
\label{Results}

\begin{figure}[tbh!]
\includegraphics{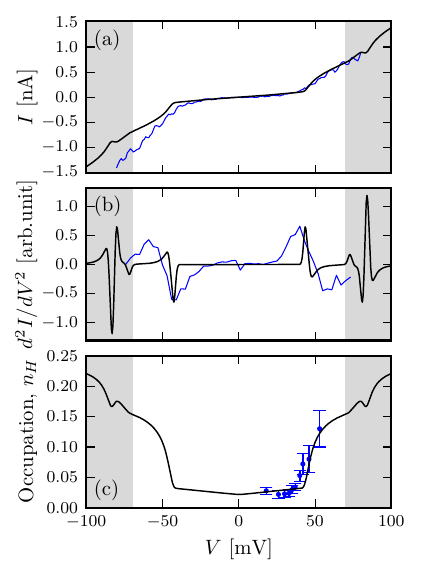}
\caption{Theoretical (black) and experimental (blue) data for (a) $I(V)$ after subtraction of measurements for clean surface, (b) IETS spectra, and (c) occupation of HC state.
\label{diffcond}}
\end{figure}

At low temperatures ($kT=0.5$ meV $\ll \hbar\Omega^L$) and voltages, direct tunneling between the two goundstates is the only possible reaction pathway. In this case, the reaction rate of the monomer is exponentially dependent on the potential energy surface (PES). The parameters used to describe the PES were therefore chosen to approximate the experimental low bias HC occupation, \Figref{diffcond}(c), and reaction rate while being close to the DFT potential, see \Tabref{table1}. The experimentally\cite{Kumagai2009} measured flip frequency of the OD monomer is $(0.9\pm0.4)\times10^3$ s$^{-1}$ at low bias (24 mV, 5 nA), and the phonon mode along the reaction coordinate (rot$_z$) has  an energy $\hbar\Omega^L=44\pm3$ meV.  The quartic polynomial used to model the PES is determined from the barrier height $V_{\text{max}}$, phonon frequency ($\hbar\Omega^L$) and asymmetry ($V_{\text{asym}}=2.0$ meV). The quartic potential shape well approximate the PES, see \Figref{PES}(b), which was obtained by fixing the substrate Cu atoms and relaxing the monomer in the direction of the surface normal. Since the fixed substrate raise the energies, a more accurate estimate of the reaction barrier was obtained using the difference in total energy of the inclined geometries and the transition state, i.e., relaxed substrate with oxygen and hydrogen fixed along the surface normal over the bridge site. This reaction barrier was found to be $V_{\text{max}}=156$ meV compared to $210$ meV for the fixed substrate. The quartic potential polynomial is then determined by $V_{\text{max}}$ and the arc length between the potential minima. Since the oxygen atom is only displaced by 0.06 \AA~during the reaction, we simplify the calculations and describe the motion using the deuterium (hydrogen) alone with a one-dimensional quartic potential. The experiment shows a fractional occupation of the HC configuration that depends on the STM tunneling current, i.e., tip-adsorbate distance. Decreasing the current attenuates the fractional occupation, which implies a decreasing asymmetry. We confirm the tip-induced asymmetry between the wells by calculating a PES at various tip-adsorbate distances. However, in the presented calculations we use a smaller tip-adsorbate distance compared to experiment in order to have a measurable overlap between the tip- and adsorbate atomic orbitals. Hence the energy difference between the wells, 2.0 meV, is chosen so that the fractional occupation of the HC state and the ground state reaction rate (24mV/5nA) match the experimental low bias values fairly well. The ground state tunneling rate and HC fractional occupation with asymmetry 2.0 meV is 40 s$^{-1}$ and 0.03 respectively. Asymmetries 1, 3 and 5 meV give 80 s$^{-1}$/0.13, 26 s$^{-1}$/0.008 and 18 s$^{-1}$/0.003 respectively. The decreasing reaction rate with increasing asymmetry is caused by the higher degree of localization of the vibrational states. 

\begin{table}[tbh!]
\caption{Comparison of computational results for the reaction barrier. The reaction rate $R(V,I)$ was calculated with our method using $V=24$ mV, $I=5$ nA, $V_{\text{asym}}=2.0$ meV.\label{table1}}
\centering
\begin{tabular}{r|rcccc}
	\toprule
	\multicolumn{2}{c}{}\\ 
	&Package&$V_{\text{max}}$&$\Delta d_{\text{min}}$& $\hbar\Omega^L$ &$R(V,I)$\\
	We &\begin{scriptsize}{SIESTA}\end{scriptsize}&156 & 2.2 & 44 & 10\\
	Ref. \citen{Kumagai2009} &\begin{scriptsize}{STATE}\end{scriptsize}\cite{Morikawa1995}& 140 & 2.1 & 43 & 220\\
	Ref. \citen{Davidson2010} &\begin{scriptsize}{VASP}\end{scriptsize}\cite{Kresse1996}& 166 & 2.1 & 46 &  20\\
	Ref. \citen{Davidson2010} &\begin{scriptsize}{CASTEP}\end{scriptsize}\cite{Clark2005}&155&2.2&43&10\\
	\hline
	Used&&150 meV& 2.15 \AA & 43 meV & 40 s$^{-1}$\\   
\end{tabular}
\end{table}

\begin{figure}[tbh!]
\begin{tikzpicture}
    \node[anchor=south west,inner sep=0] at (0,0) {\includegraphics{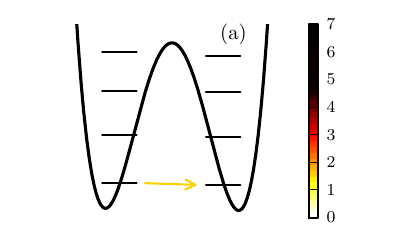}};
    \draw[ thin,stealth-stealth] (1.75,.15)--(4.05,.15);
    \node at (3,.35) {2.15\AA};
    \draw[densely dotted] (0.95,.5)--(4.05,.5);
    \draw[densely dotted] (0.95,.55)--(1.75,.55);
    \draw[ thin,-stealth] (0.958,.75)--(0.958,.55);
    \draw[ thin,-stealth] (0.958,.31)--(0.958,.5);
    \node at (0.2,.515) {2.0 meV};
    \draw[densely dotted] (0.95,.99)--(1.75,.99);
    \draw[densely dotted] (0.95,1.81)--(1.75,1.81);
    \draw[densely dotted] (0.95,2.56)--(1.75,2.56);
    \draw[ thin,stealth-stealth] (.95,.99)--(.95,1.81) ;    
    \draw[ thin,stealth-stealth] (.95,1.81)--(.95,2.56) ;    
    \node at (0.2,1.45) {44 meV};
    \node at (0.2,2.20) {39 meV};    
    \node at (0.2,3.60) {$V=24$ meV};
    \node at (0.05,3.2) {$I=5$ nA};   

    \node[anchor=south west,inner sep=0] at (0,-4) {\includegraphics{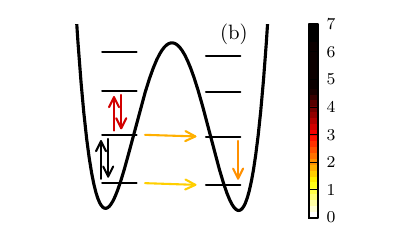}};
    \node at (0.2,-.4) {$V=64$ meV}; 
    \node at (0.05,-.8) {$I=5$ nA};   
    \node[anchor=south west,inner sep=0] at (0,-8) {\includegraphics{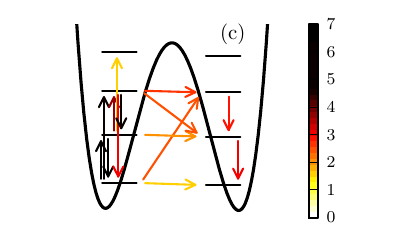}};
    \node at (0.2,-4.4) {$V=94$ meV}; 
    \node at (0.05,-4.8) {$I=5$ nA};   
    \draw[ thick,-stealth] (1.7,-8)--(4.0,-8) ; 
    \node at (2.8,-7.8) {\tiny{Reaction coordinate}};      
\end{tikzpicture}
	\caption{PES for the OD rot$_z$ mode including a logarithmic scaled (base 10) transition rates [s$^{-1}$] and path of reaction at 5 nA tunneling current for various bias voltages, illustrating the importance of ladder climbing and multi-phonon emission. A lower bound cutoff ($30$ s$^{-1}$) is used to highlight the important paths.\label{PathFig}}		
\end{figure}
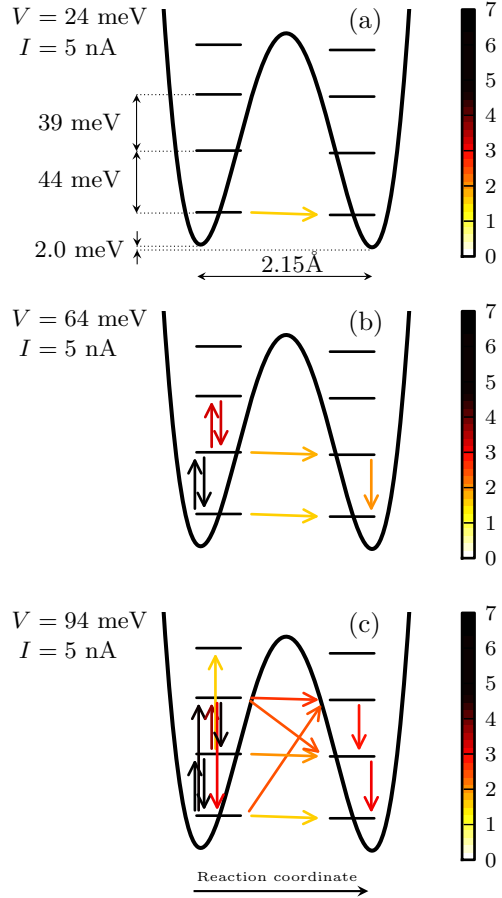

To provide a physical picture of the flipping events, the reaction pathways from the left to right potential wells are shown in \Figref{PathFig}. The reaction proceed by direct tunneling between the ground states at low bias $V=24$ meV, \Figref{PathFig}(a). For the parameters used here, the reaction rate $~40$ s$^{-1}$ is slightly lower than the experimentally measured reaction rate. However, the sensitivity of the rate to the exact parameters used, indicate that we are only able to model the reaction rate qualitatively. As the bias is increased above the fundamental frequency  ($V=64$ meV), vibrationally assisted tunneling becomes the most important pathway, \Figref{PathFig}(b). Although emission of vibrational quanta increase the population of excited states in the left well, the large \textit{e-h} pair damping rate prevents the ladder climbing and reactions occurring by going over the barrier. However, in the excited states, the effective barrier becomes lower and narrower and the reaction rate increases significantly. 

Increasing the bias voltage further ($V=94$ mV) increase the reaction rate and processes involving double excitations, due to the anharmonic PES, become energetically allowed, \Figref{PathFig}(c). Here, the direct transition from $L_0$ to $R_2$ contribute together with VAT from $L_2$ to $R_2$ where $L_2$ is populated from the left ground state by direct excitation. Ladder climbing processes are visible although relatively slow due to the large \textit{e-h} pair damping rate compared to the excitation rate calculated for the rot$_z$ mode. The ratio between overtone excitation- and fundamental excitation rate in the left well is only $\gamma_{0\to2}/\gamma_{0\to1}\approx 1\%$, however the tunneling barrier between $L_2 $ and $R_2$ is significantly smaller than the barrier between $L_1$ and $R_1$, explaining why the double excitation dominates the reaction at this bias voltage.

\begin{figure}[tbh!]
	\begin{tikzpicture}
		\node at (.705,0) {\includegraphics{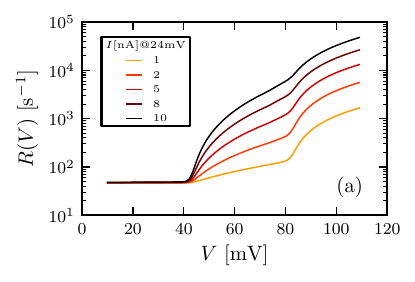}};
		\node at (.6,-4.5) {\includegraphics{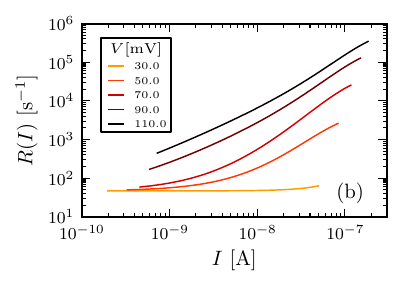}};
	\end{tikzpicture}
	\caption{(a) Reaction rate versus bias voltage for various tunneling currents, and (b) reaction rate versus current for different bias voltages\label{rate}}		
\end{figure}

The current-voltage characteristics of the STM experiment is indirectly determined by the reaction rates shown in \Figref{PathFig}. The time averaged current-voltage characteristics is determined by the steady state occupation of the left/right states which in turn is determined by the ratio of the left/right transition rates.  The $I(V)$ curve, \Figref{diffcond}(a), is calculated from \Eqref{d2idv2} where a background is subtracted to account for clean surface measurements, as they do in the experiment. The IETS spectra, \Figref{diffcond}(b), is calculated as the second derivative of $I(V)$, and the calculated HC occupation is shown in \Figref{diffcond}(c) together with experimental data. The IETS spectra shows signals at $\pm$43 ($L_0 \to L_1$) and $\pm 83$ ($L_0 \to L_2$) meV with a smaller signal at $\pm 72$ ($L_1 \to L_3$) meV. Note that the IETS spectra considered here is simply an effect of the difference in conductance between the left/right states. The conventional IETS directly caused by the inelastic scattering of the electrons is not considered here since it is in general much weaker than the large conductance change of the flipping motion. The vibrational energies clearly deviate from the harmonic approximation due to the anharmonicity of the PES, i.e., the vibrational energies are not integer multiplies of the fundamental frequency.  The corresponding experimental signals show up at 41 (-45) meV. However, the bias region above 70 mV, roughly the shaded regions in \Figref{diffcond}, is not included in the experimental HC occupation and IETS, hence we do not have evidence of possible overtones, and thereby our predicted anharmonicity of the PES. According to our theory the overtones should give signals of similar strength around the overtone frequency. However, in our model, the magnitude of the overtone signal is sensitive to the computational parameters, especially the current level, since it is caused by the rapid oscillation of the occupations and not an overall shift. This is due to the fact that the higher excited states have much larger left/right transition rates, although, as explained above, the overtone excitation itself is suppressed compared to first order excitations. In addition, higher order processes can be expected to be significant at this bias since second order processes are not symmetry forbidden for the harmonic oscillator, e.g., the $L_0 \to L_2$ transition. We therefore recommend caution in respect to the analysis of results above 70 mV.

The reaction rate, $R(V)$, at a fixed tip-sample separation as a function of voltage is shown in \Figref{rate}(a) where the tip-sample separation was determined by fixing the current (at 24 mV). The reaction rate is nearly constant below the vibrational energy due to direct tunneling between the ground states. The onset of inelastic excitations by the tunneling electrons is clearly seen at the fundamental excitation $\hbar\Omega$ and overtone excitation $\hbar\Omega_{02}$, i.e., at the same frequencies as the signals in the IETS curve. The reaction rate $R(I)$ as a function of the current at fixed bias is shown in \Figref{rate}(b). At low bias, the rate is not current dependent since the direct tunneling reaction pathway is not affected by the tunneling electrons. At moderate bias $\hbar\Omega<V<\hbar\Omega_{02}$ and currents above $10$ nA, the main reaction pathway is the two electron process, i.e., ladder climbing to the second excited state before tunneling. The current dependence of the reaction rate is therefore approximately $R(I)\propto I^2$.\cite{Stipe1997} At even higher bias, one electron processes overtake direct tunneling and ladder climbing resulting in a reaction rate directly proportional to the current. 

In summary, we have developed a new method to calculate the tunneling electron induced reaction pathways and rates for adsorbates on surfaces using first principles methods. For the hydroxyl species on a Cu(110) surface, the reaction rate of the back and forth flipping motion is well described by the method while also giving insights into the reaction pathways. The large \textit{e-h} pair damping rate quenches ladder climbing and makes  multi-phonon transitions pronounced in the differential conductance plot at large biases. This crossover underlines the influence of the anharmonicity of the underlying potential energy surface.  The newly developed method should be extendable to more complicated adsorbates opening up for a more detailed understanding of reactions on surfaces, including energy transfer between different vibrational states.

\section{acknowledgement}
The computations were performed on resources provided by the Swedish National Infrastructure for Computing (SNIC) at Lunarc. A.G. and M.P. are supported by a grant from the Swedish Research Council (621-2010-3762).
H.U. is supported by a Grant-in-Aid for Scientific Research C (No. 25390007) from JSPS. We thank T. Kumagai for sharing the experimental data.

\bibliographystyle{apsrev}

\end{document}